\documentclass[a4paper,11pt]{article}
\usepackage{pos}
\usepackage{physics}
\usepackage[capitalize]{cleveref}
\usepackage{subcaption}
\usepackage{wrapfig}
\usepackage{bigints}

\usepackage{hyperref}

\newcommand{\xpom}{x_{I\hspace{-0.3em}P}}
\newcommand{\xbj}{{x}}
\newcommand{\ygap}{Y_{\rm gap}}

\newcommand{\gev}{\mathrm{GeV}}
\newcommand{\Pev}{\mathrm{PeV}}

\newcommand{\Tev}{\mathrm{TeV}}

\newcommand{\rt}{{\mathbf{r}}}

\newcommand{\bt}{{\mathbf{b}}}

\title{High-energy nuclear scattering of neutrinos}
\author*[a,b]{Anh Dung Le}
\author[a,b]{Heikki Mäntyssari}

\affiliation[a]{Department of Physics, University of Jyväskylä,\\
  P.O. Box 35, 40014 University of Jyväskylä, Finland}

\affiliation[b]{Helsinki Institute of Physics,\\
 P.O. Box 64, 00014 University of Helsinki, Finland}

\emailAdd{anh.d.le@jyu.fi}
\emailAdd{heikki.mantyssari@jyu.fi}

\abstract{
We study the energy dependence of the total and diffractive neutrino-nucleon and neutrino-nucleus cross sections at very high energies. The calculation employs the QCD dipole model and the small-$\xbj$ nonlinear Balitsky-Kovchegov evolution. We show the sensitivity of the nuclear effect quantification on the nuclear setup, and predict up to $\sim 10\%$ nuclear suppression in the inclusive neutrino-oxygen scattering stemming from the nonlinear evolution. Diffractive contribution to the total scattering is small, which is only few percentage. The $\ket{q\bar{q}g}$ componnent of the $W^{\pm}$ boson is found to contribute significantly to the diffractive process, which reaches up to $\sim 40\%$ of the diffractive cross section.
}

\FullConference{%
  42nd International Conference on High Energy Physics (ICHEP2024)\\
  18-24 July 2024\\
  Prague, Czech Republic
}

\begin{document}
\maketitle

\section{Introduction}
\label{sec:intro}

Over the years, various detectors and telescopes have been constructed or designed in order to unveil elegant puzzles of the ghost particles - neutrinos~\cite{SajjadAthar:2021prg}. Many facilities are designed toward observing neutrinos in the high-energy (HE, $1, \Tev \le E_\nu \le 100\,\Pev$) and ultra-high-energy (UHE, $E_\nu \ge 100 \, \Pev$) regimes. Since neutrinos cannot be detected directly, but only through the secondary outcomes from their weak interactions with nucleons or nuclei, understanding of their scattering behaviors with these targets is highly demanded.

In the HE and UHE domains of interest, the interaction of neutrinos with nucleons and nuclei is dominated by deep inelastic scattering. The small Bjorken-$\xbj$ dynamics becomes important in the description of the scattering process as such energies. Involving the small-$\xbj$ description in deep-elastic scattering processes can be conveniently done by incorporating the QCD dipole model~\cite{Mueller:1989st,Nikolaev:1990ja} and a high-energy evolution equation. Such an approach has been employed in e.g Refs.~\cite{Kutak:2003bd,Albacete:2015zra,Ducati:2006vh} to investigate the inclusive scattering of neutrinos off nucleons and nuclei.

In the current work, we employ the QCD dipole model and the nonlinear Balitsky-Kovchegov (BK) equation~\cite{Balitsky:1995ub,Kovchegov:1999yj,Balitsky:2006wa} at small Bjorken-$\xbj$ to study inclusive and diffractive scattering of neutrinos off both nucleons and nuclei (in particular, oxygen). The latter equation governs the energy dependence of the dipole-target scattering amplitude, which is the basic ingredient of the dipole model. The nonlinear nuclear modification effect is quantified by constructing the dipole-nucleus amplitude from the dipole-nucleon amplitude using the optical Glauber model at two stages, either before or after the small-$\xbj$ evolution.

\section{Dipole formulation for neutrino-nucleon and neutrino-nucleus scatterings}
\label{sec:formulation}
In the limit of massless quarks, the cross section for the total and diffractive  neutrino-nucleon (nucleus) scatterings by the exchange of a vector boson ($W^{\pm}/Z$) of virtuality $Q^2$ reads
\begin{equation}
\label{eq:xsections}
    \frac{\dd \sigma_{\nu N(A);{\rm tot/D}}^{W^{\pm}/Z}}{\dd\mathfrak{Inv}_{\rm tot/D}} = \frac{G_F^2}{4\pi x} \left(\frac{M_{W^{\pm}/Z}^2}{M_{W^{\pm}/Z}^2 + Q^2}\right)^2 \left\{ \left[1 + (1-y)^2\right] F_{2;N(A)}^{{\rm tot/D(3)};W^{\pm}/Z} - y^2F_{L;N(A)}^{{\rm tot/D(3)};W^{\pm}/Z} \right\},
\end{equation}
where $G_F$ is the Fermi constant. The inelasticity $y$ is related to the Bjorken-$\xbj$ variable and $Q^2$ as $y=Q^2/(xs)$, where $s=2m_N E_\nu$ is the squared center-of-mass energy and $E_\nu$ is the neutrino energy. Inclusive scattering kinematics can be described by two invariants $(\xbj, Q^2)$, while the diffractive process depends additionally on $\xpom$ controlling the rapidity gap, $\ygap = \ln (1/\xpom)$, within the total rapidity interval $Y = \ln(1/\xbj)$. Consequently, $\dd\mathfrak{Inv}_{\rm tot} \equiv \dd\xbj \dd Q^2$ and $\dd\mathfrak{Inv}_{\rm D} \equiv \dd\xpom\dd\xbj \dd Q^2$.   

Both inclusive and diffractive scattering processes can be described using the QCD dipole model. In this picture, at the leading order, the inclusive structure functions $F_{T,L;N(A)}^{{\rm tot};W^{\pm}/Z}$ for transverse (T) and longitudinal (L) polarizations can be written in terms of the forward  elastic dipole-target amplitude $N(\xbj,\rt,\bt)$ as
\begin{equation}
\label{eq:dip_factorization_incl}
    F_{T,L;N(A)}^{{\rm tot};W^{\pm}/Z} = \frac{Q^2}{4\pi \alpha_{W/Z}} \int \dd^2 \rt \int_0^1 \dd z \left|\Psi_{T,L}^{W^{\pm}/Z\to q\bar q}(\rt,z,Q^2)\right|^2 \int \dd^2\bt~ 2N(\xbj,\rt,\bt),
\end{equation}
where $\alpha_W = \frac{e^2}{32\pi \sin^2\theta_W}$ and $\alpha_Z = \frac{e^2}{16\pi \sin^2\theta_W\cos^2 \theta_W}$,
with $\theta_W$ being the Weinberg's angle. The squared wave functions $\left|\Psi_{T,L}^{W^{\pm}/Z\to q\bar q}\right|^2$ can be found e.g. in Refs.~\cite{Albacete:2015zra,Le:2024wvy}. 

In the CGC framework, and in the mean-field approximation, the dipole-target amplitude can be obtained by solving numerically the nonlinear Balitsky-Kovchegov (BK) equations~\cite{Balitsky:1995ub,Kovchegov:1999yj}. We choose to solve the leading-order equation with running coupling correction encoded in the Balitsky's integration kernel \cite{Balitsky:2006wa}. 
Instead of solving the full $\bt$-dependent equation, we assume that in the case of a nucleon target the $\bt$-dependence can be factorized as $N(\xbj,\rt,\bt) = T_N(\bt)\mathcal{N}(x,\rt)$, with the density profile $T_N$ normalized as $\int \dd^2\bt T_N(\bt) = \sigma_0 /2$.
The amplitude $\mathcal{N}(x,\rt)$ solves the $\bt$-independent BK with an initial condition parametrized at $x=x_0=0.01$ given in Refs.~\cite{Lappi:2013zma,Lappi:2023frf}.

The free parameters in the setup can be determined by fitting to the HERA inclusive cross section data. Here we use the two light-quark fits from Ref.~\cite{Lappi:2013zma} referred to as ${\rm MV}$ and ${\rm MV}^e$. 

Generalization to nuclear targets can be done via two different scenarios. First, the dipole-nucleus amplitude at each impact parameter evolves independently from each other by the impact-parameter-independent BK equation. The initial condition for this evolution is constructed from the initial dipole-nucleon amplitude at $x=\xbj_0$ using the optical Glauber model~\cite{Lappi:2013zma,Kowalski:2003hm} by employing an independent scattering approximation 
\begin{equation}
\label{eq:nuclear-amplitude}
    N_A(\xbj_0,\rt,\bt) = 1 - \left[1-\frac{\sigma_0}{2}T_A(\bt)\mathcal{N}(x_0,\rt)\right]^A,
\end{equation}
where $T_A(\bt)$ is the Woods-Saxon density. This setup is referred to as {\em before evol}. In order to quantify the saturation effect in the evolution, we also consider the so-called {\em after evol.} setup, in which the dipole-nucleus amplitude $N_A(\xbj,\rt,\bt)$ is obtained from the BK-evolved dipole-nucleon amplitude $\mathcal{N}(\xbj,\rt)$ by using~\cref{eq:nuclear-amplitude} at the desired $\xbj$.

The coherent diffractive production can be seen as the quasi-elastic scattering of the vector boson's Fock state off the target. In this sense, the dipole factorization for the diffractive structure functions $F_{(x)T,L;\nu N(A)}^{D(3);W^{\pm}/Z}$ of the Fock state $\ket{x}$ schematically reads
\begin{equation}
\label{eq:dfrac_sf}
    F_{(x)T,L;\nu N(A)}^{D(3);W^{\pm}/Z} \propto \frac{Q^2}{4\pi \alpha_{W/Z}} \frac{Q^2\xpom}{\xbj} \sum_{\mathrm{pol.}} \left[\left(\Psi^{W^{\pm}/Z\to x}_{T,L;(C)}\right)^{*} \Psi^{W^{\pm}/Z\to x}_{T,L}\right]\otimes \left[\mathfrak{N}_{xN(A);(C)}\mathfrak{N}_{xN(A)}\right],
\end{equation}
where $\Psi^{V\to x}_{T,L}$ are the wave functions describing the fluctuation $V\to x$ for the vector boson $V$ with transverse (T) or longitudinal (L) polarizations,
 and $\mathfrak{N}_{xA(N)}$ is the amplitude for the scattering of the Fock state $\ket{x}$ off the nuclear (nucleon) target. The subscript $``(C)"$ refers to the quantities in the complex-conjugated amplitude. The sum over possible polarizations and helicities (pol.) of the $\ket{x}$ state is done. 
 We consider only the two lowest-order states $\ket{q\bar{q}}$ and $\ket{q\bar{q} g}$, which are expected to provide the most important contributions at large-$Q^2$. The corresponding diffractive structure functions are given in Ref.~\cite{Le:2024wvy}. Note that they are sensitive to the density $T_N$ in the case of a nucleon target, for which we  use the incomplete gamma function profile specified in  Ref.~\cite{Lappi:2023frf}. In addition, for the $\ket{q\bar{q}g}$ component at large-$Q^2$, only the transverse polarization is relevant. 

We shall focus on the energy dependence of the integrated total/diffractive cross sections defined by
\begin{equation}
    \sigma_{\nu N(A);\rm tot/D}^{W^{\pm}/Z}(E_{\nu}) = \bigintsss_{\begin{smallmatrix}
1~\gev^2 \le Q^2 \le \xbj_{\rm max}s \\
Q^2/s\le \xbj \le \xbj_{\rm max}\\
\xbj \le \xpom \le \xbj_{\rm max}
\end{smallmatrix}}\dd \mathfrak{Inv}_{\rm tot(D)} \frac{\dd \sigma^{W^{\pm}/Z}_{\nu N(A);\rm tot(D)}}{\dd \mathfrak{Inv}_{\rm tot(D)}}.
\end{equation}

Here the value of the longitudinal cut-off $x_{\rm max}$ defines the scheme used to large-$x$ extrapolation. If $\xbj_{\rm max} = \xbj_{0}$, then only the small-$\xbj$ contribution is included in the calculations. Otherwise, if $\xbj_{\rm max} = 1 $, we extrapolate to large-$\xbj$ regime as
\begin{equation}
    \label{eq:large-x-amp}
    \mathcal{N}(\xbj,\rt) = \mathcal{N}(\xbj_0,\rt) \left(\frac{1-\xbj}{1-\xbj_0} \right)^{6}.
\end{equation}

\section{Results}
\label{sec:results}
\begin{figure*}[t!]
\label{fig:inclusive_xsec}
    \centering
    \begin{subfigure}[t]{0.49\textwidth}
        \centering
        \includegraphics[width=\textwidth]{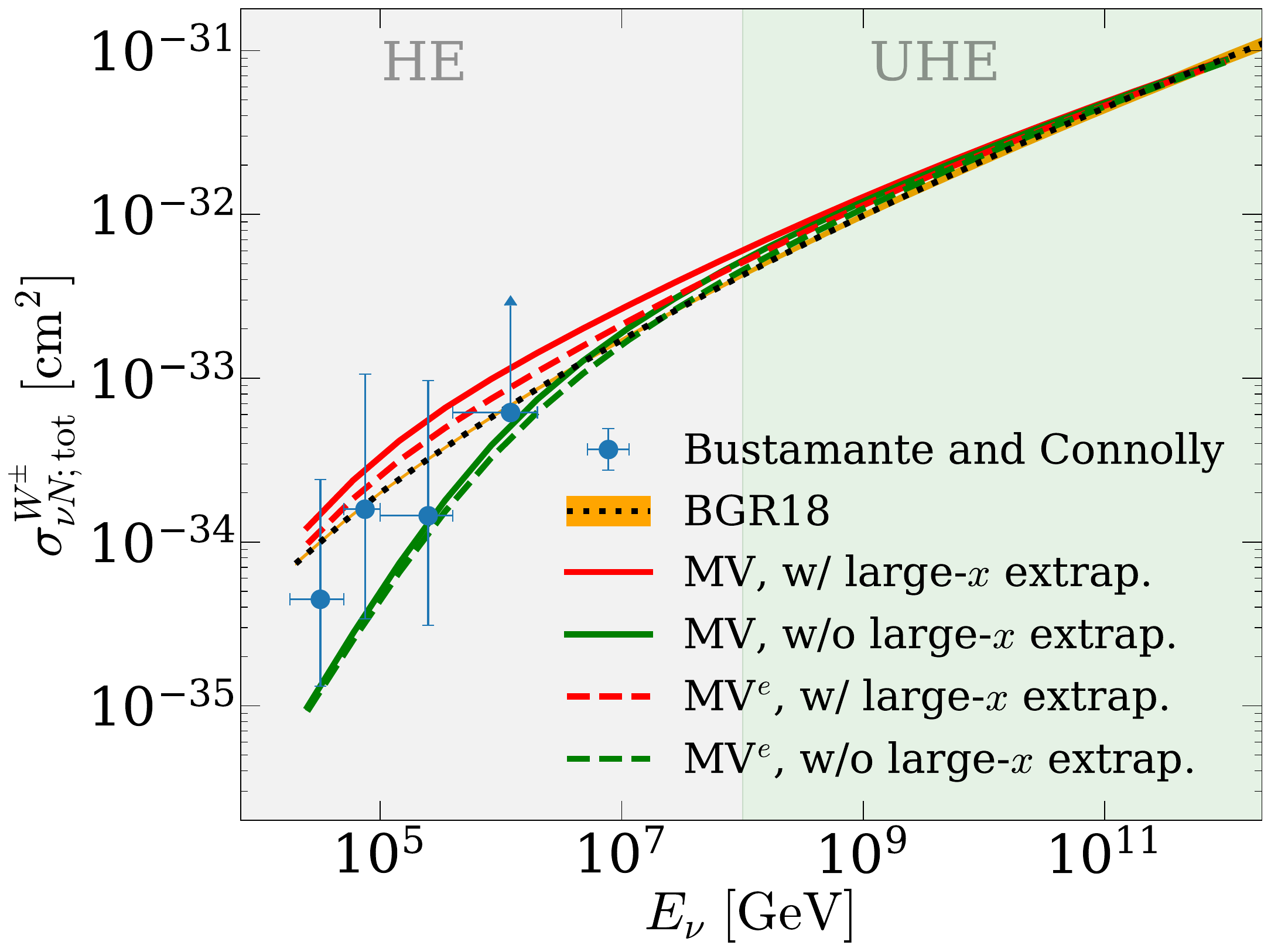}
        \caption{Nucleon target}
        \label{subfig:incl_nucleon}
    \end{subfigure}%
    ~ 
    \begin{subfigure}[t]{0.49\textwidth}
        \centering
        \includegraphics[width=\textwidth]{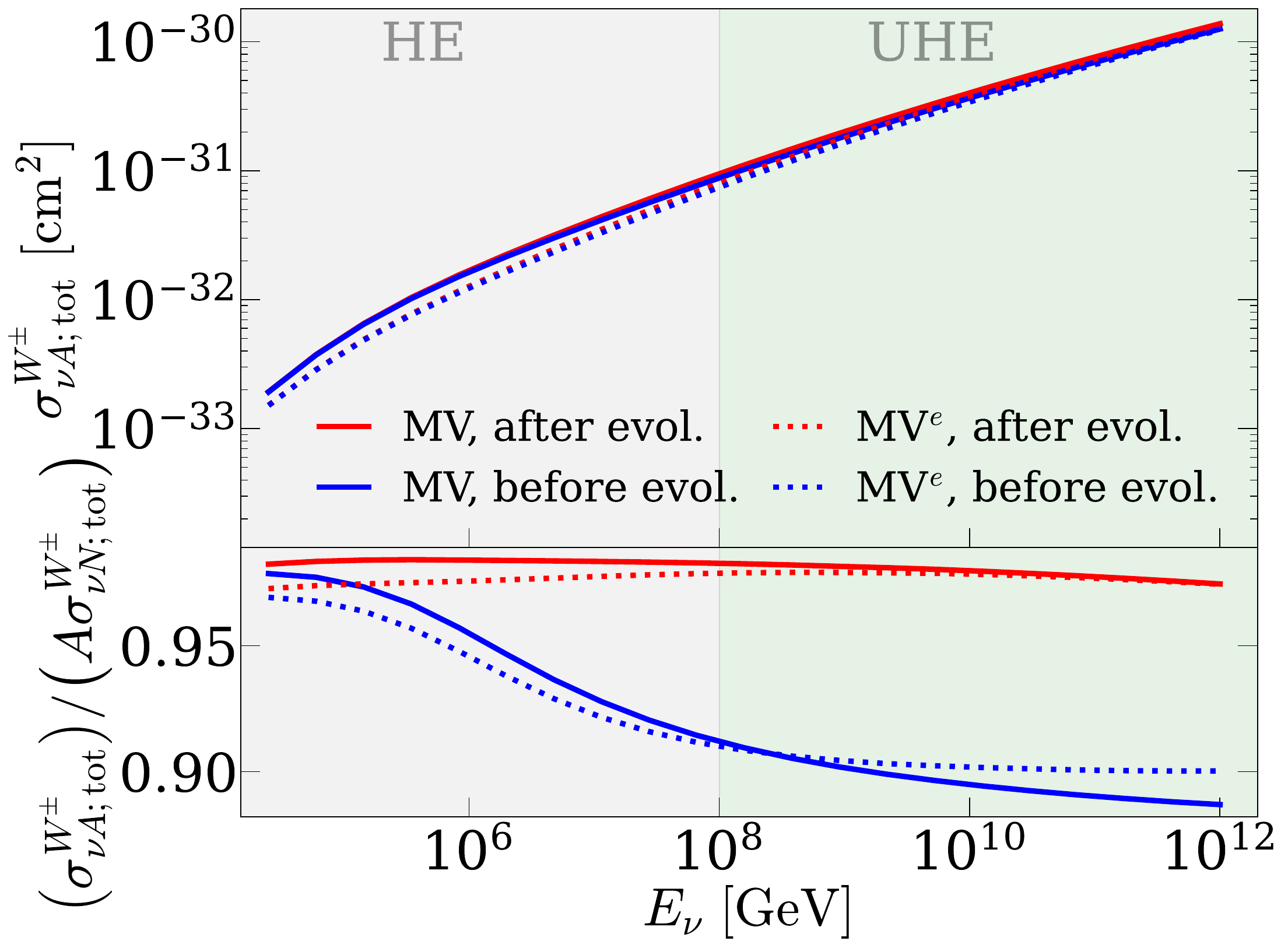}
        \caption{Nuclear target (Oxygen $^{16}O$)}
        \label{subfig:incl_nucleus}
    \end{subfigure}
    \caption{Inclusive neutrino-nucleon (\subref{subfig:incl_nucleon}) and neutrino-oxygen (\subref{subfig:incl_nucleus}) cross-sections for charged current ($W^{\pm}$) exchange. The ``BGR18'' curve represents the result from Ref.~\cite{Bertone:2018dse}. The blue points are extracted from the IceCube data in Ref.~\cite{Bustamante:2017xuy}. 
    The nuclear modification ratio is shown in the second row of (\subref{subfig:incl_nucleus}). Results for the nucleon target are shown both with and without the large-$x$ extrapolation, while only results with large-$\xbj$ extrapolation are shown for the nuclear target.}  
\end{figure*}

The total (inclusive) neutrino-nucleon cross section for
neutrino energies covering the high-energy (HE) and
ultra-high-energy (UHE) regimes are shown in \cref{subfig:incl_nucleon}. 
The cross section is found to depend strongly on the  large-$\xbj$ extrapolation at neutrino energies below $\sim 10^{7} ~\gev$. However, within the uncertainty, the results from both schemes are compatible with the available IceCube
data in this domain extracted in Ref.~\cite{Bustamante:2017xuy}. Meanwhile, the cross sections for the UHE neutrinos are shown to be sensitive only to the low-$\xbj$ regime. 

The obtained cross sections with large-$\xbj$ extrapolation are close to the results from the collinear factorization approach at NNLO including the resummation of small-$\xbj$ linear BFKL effects reported in Ref.~\cite{Bertone:2018dse} (BGR18). This is especially true when the ${\rm MV}^e$ fit is applied.
 However a slight suppression in the rise of the cross sections calculated using the CGC approach of the current
work compared to BGR18 can be seen for energies above $\sim 10^{11} ~\gev$. This observation suggests that the gluon saturation effect can become relevant at such energies. In addition, one can see that the ${\rm MV}^e$ fit provides a better description of the extracted IceCube data. The difference between the MV and the ${\rm MV}^e$ fits becomes negligible in the UHE domain.     

A similar energy dependence to the nucleon case is observed for the total neutrino-oxygen cross sections, as shown in \cref{subfig:incl_nucleus}. Again, the ${\rm MV}^{e}$ initial condition provides a slightly smaller prediction for the cross section than the $\rm{MV}$ fit. The difference between two methods chosen to generalize the dipole-nucleon amplitude to the dipole-nucleus
case is not visible in the cross sections. However, it becomes obvious in the nuclear modification ratio shown in the lower panel of \cref{subfig:incl_nucleus}. In the {\em after evol.} setup, the nuclear effect is small (about $2\%$ suppression) and the ratio is almost energy-independent. Meanwhile, an obvious energy dependence  is visible in the nuclear modification factors calculated using the {\em before evol.} setup. The nuclear suppression in this setup is about $2\hdots 5\%$ in the HE region covered by current available data, and up to $\sim 10\%$ in the UHE regime. This scheme dependence shows that the total neutrino-nucleus cross section is sensitive to the saturation effects in the small-$\xbj$ evolution, as in the {\em before evol.} the nonlinear effects are enhanced in the evolution due to the large saturation scale at the initial condition compared to the  {\em after evol.} scheme.

\begin{figure*}[t!]
\label{fig:diffractive_xsec}
    \centering
    \begin{subfigure}[t]{0.49\textwidth}
        \centering
        \includegraphics[width=\textwidth]{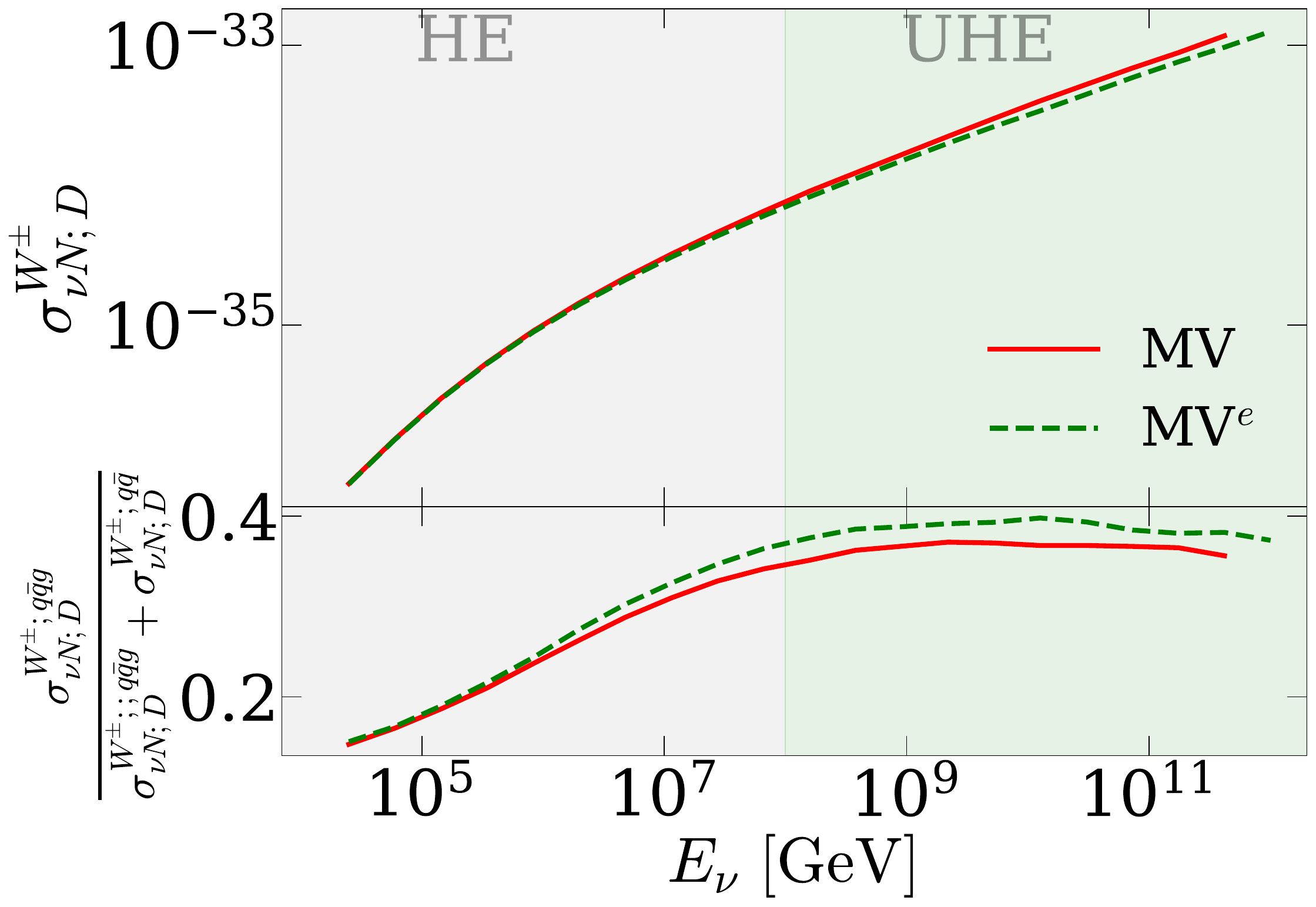}
        \caption{Nucleon target}
        \label{subfig:dfrac_nucleon}
    \end{subfigure}%
    ~ 
    \begin{subfigure}[t]{0.49\textwidth}
        \centering
        \includegraphics[width=\textwidth]{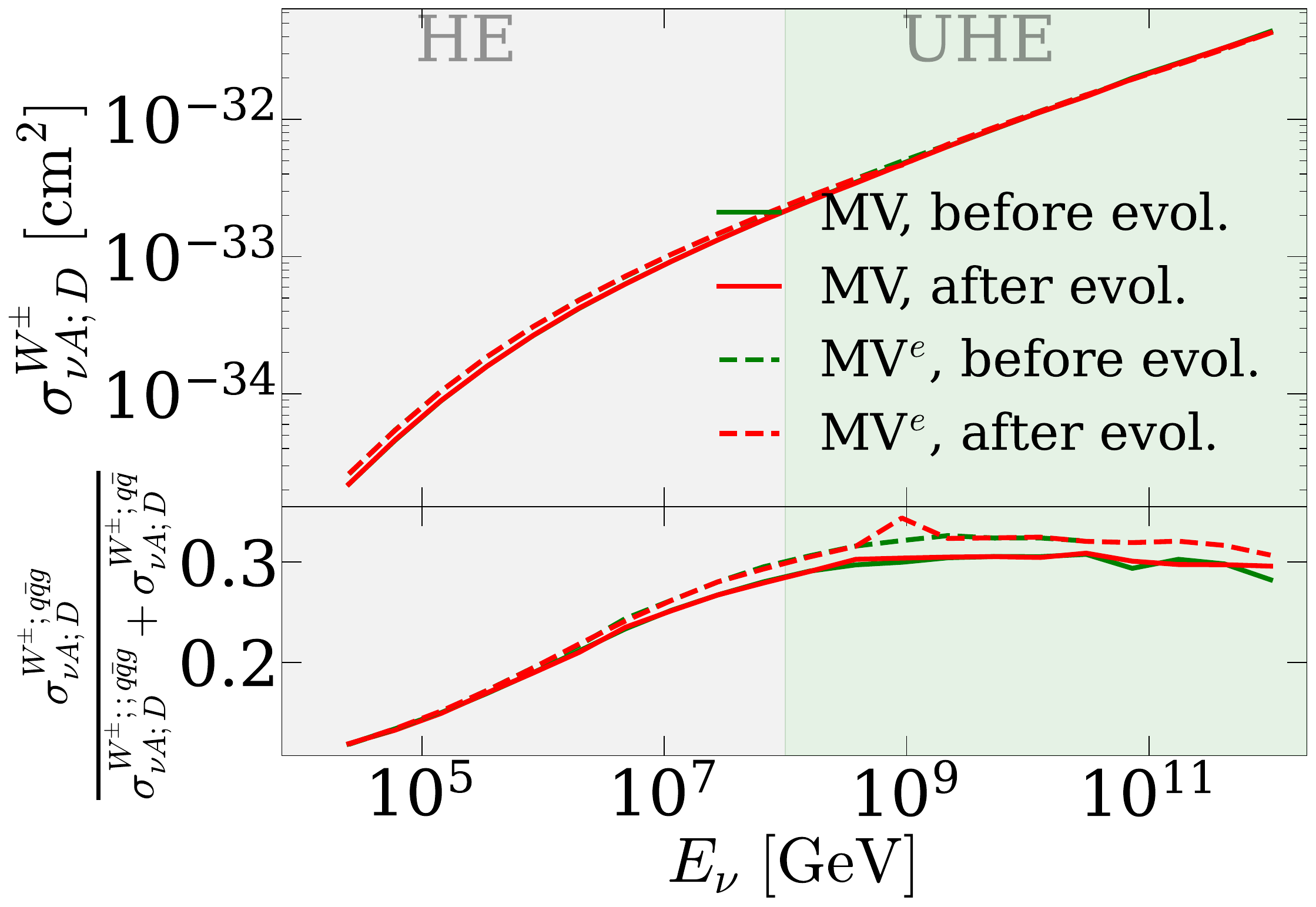}
        \caption{Nuclear target (Oxygen $^{16}O$)}
        \label{subfig:dfrac_nucleus}
    \end{subfigure}
    \caption{Coherently diffractive neutrino-nucleon (\subref{subfig:dfrac_nucleon}) and neutrino-oxygen (\subref{subfig:dfrac_nucleus}) cross sections for charged current ($W^{\pm}$) exchange. The second row shows the fraction of the $\ket{q\bar{q}g}$ contribution to the cross sections. Only results with large-$\xbj$ extrapolation are shown for both targets.}
\end{figure*}

The coherently diffractive cross sections for the neutrino-nucleon and neutrino-oxygen scattering processes are presented in \cref{subfig:dfrac_nucleon} and \cref{subfig:dfrac_nucleus}, respectively. Diffractive contribution in the total cross section is found to be small, only about $1\%$ for the nucleon target and few percent for the oxygen target. As for the inclusive scattering, similar energy-dependent behavior is observed for both nucleon and oxygen targets. In this case, the dependency on the setup is weak. The contribution from the $\ket{q\bar{q}g}$ component increases
with increasing neutrino energy until a plateau is reached
in the UHE regime. This behavior comes from the interplay between two effects when the energy increases: the enlargement of the available phase space for the gluon contribution, and the black disc limit. The former is dominantly important in the HE regime, while the latter starts to play the role when entering the UHE domain. Furthermore, the relative gluon contribution is slightly smaller in the scattering off oxygen, which is also a consequence of the black dics limit. 

More detailed discussions together with the results for the neutral-current exchange and the incoherently diffractive production can be found in Ref.~\cite{Le:2024wvy}.

\section{Conclusions}
\label{sec:conclusions}
We have presented the predictions for both inclusive and diffractive neutrino-nucleon and neutrino-nucleus cross sections by the exchange of $W^{\pm}$ bosons in the QCD dipole picture at very high neutrino energies $E_{\nu} > 10 ~\Tev$. The energy dependence is extracted by solving the small-$\xbj$ nonlinear BK evolution, with all inputs being constrained using the HERA electron-proton data. This calculation is shown to provide a good description to the currently available data of total cross section in the HE regime. The total neutrino-nucleus cross section is sensitive to the gluon saturation, as is manifested in the dissimilarity in the nuclear modification factor between two ``nucleation" schemes, {\em after evol.} and {\em before evol.}. The latter case can result in up to $\sim10\%$ suppression for the UHE neutrinos. The involved $\ket{q\bar{q}g}$ component is found to be important for the diffractive process, which can account for up to $\sim 30\hdots40\%$ of the diffractive cross sections.        

\small
\begin{acknowledgments}
This work was supported by the Research Council of Finland, the Centre of Excellence in Quark Matter (project 346324 and 364191) and projects 338263, 346567 and 359902, and by the European Research Council (ERC, grant agreements ERC-2023-101123801 GlueSatLight and ERC-2018-ADG-835105 YoctoLHC). 
Computing resources from CSC – IT Center for Science in Espoo, Finland were used in this work.
The content of this article does not reflect the official opinion of the European Union and responsibility for the information and views expressed therein lies entirely with the authors. 

\end{acknowledgments}

\footnotesize
\setlength{\bibsep}{0pt plus 0.3ex}

\bibliography{refs}
%\end{thebibliography}

\end{document}